\documentclass[11pt]{article}

\usepackage[margin=1in]{geometry}
\usepackage{amsmath,amssymb}
\usepackage{graphicx}
\usepackage{hyperref}
\usepackage{natbib}
\usepackage{booktabs}
\usepackage{listings}
\usepackage{xcolor}
\definecolor{codegreen}{rgb}{0,0.6,0}
\definecolor{codegray}{rgb}{0.5,0.5,0.5}
\definecolor{codepurple}{rgb}{0.58,0,0.82}
\definecolor{backcolour}{rgb}{0.97,0.97,0.97}

\lstdefinestyle{pythonstyle}{
    backgroundcolor=\color{backcolour},
    commentstyle=\color{codegreen},
    keywordstyle=\color{codepurple},
    numberstyle=\tiny\color{codegray},
    stringstyle=\color{codegreen},
    basicstyle=\ttfamily\small,
    breakatwhitespace=false,
    breaklines=true,
    captionpos=b,
    keepspaces=true,
    showspaces=false,
    showstringspaces=false,
    showtabs=false,
    tabsize=2,
    frame=single,
    framerule=0pt,
}
\newcommand{\pkg}[1]{\texttt{#1}}
\newcommand{\proglang}[1]{\textsf{#1}}
\newcommand{\code}[1]{\texttt{#1}}
\newcommand{\fct}[1]{\texttt{#1()}}

\title{\pkg{causalfe}: Causal Forests with Fixed Effects in \proglang{Python}}

\author{Harry Aytug\\[0.5em]
\small Amazon Web Services\\
\small \texttt{haytug@amazon.com}}
\date{\today \\ \vspace{0.5em} \small Working Paper}

\begin{document}

\maketitle

\begin{abstract}
The \pkg{causalfe} package provides a \proglang{Python} implementation of 
Causal Forests with Fixed Effects (CFFE) for estimating heterogeneous 
treatment effects in panel data settings. Standard causal forest methods 
struggle with panel data because unit and time fixed effects induce spurious 
heterogeneity in treatment effect estimates. The CFFE approach addresses this 
by performing node-level residualization during tree construction, removing 
fixed effects within each candidate split rather than globally. This paper 
describes the methodology, documents the software interface, and demonstrates 
the package through simulation studies that validate the estimator's 
performance under various data generating processes.
\end{abstract}

\noindent\textbf{Keywords:} causal inference, causal forest, fixed effects, panel data, 
heterogeneous treatment effects, difference-in-differences, \proglang{Python}

\vspace{1em}


\section{Introduction} \label{sec:intro}

Panel data and difference-in-differences (DiD) designs have become central to 
empirical research in economics and the social sciences. These settings often 
involve unit-specific and time-specific unobserved heterogeneity that must be 
controlled through fixed effects. At the same time, researchers increasingly 
recognize that treatment effects may vary across individuals---the conditional 
average treatment effect (CATE), denoted $\tau(x)$, captures this heterogeneity 
as a function of observed covariates $x$.

Causal forests \citep{Athey+Imbens:2016, Wager+Athey:2018} offer a flexible, 
nonparametric approach to estimating heterogeneous treatment effects. The method 
recursively partitions the covariate space to identify subgroups with differing 
treatment responses, then estimates local treatment effects within each leaf 
using honest sample splitting. This approach has proven effective in 
cross-sectional settings where observations are independent.

Panel data, however, present a challenge. When unit and time fixed effects are 
present, standard causal forests can produce misleading results. The problem 
arises because the forest may split on variables correlated with the fixed 
effects themselves, creating spurious heterogeneity in the estimated treatment 
effects. A unit with a high fixed effect, for instance, might appear to have a 
larger treatment response simply because its baseline outcome level is elevated. 
Global residualization---removing fixed effects before fitting the forest---does 
not fully resolve this issue, as the within-node treatment effect estimates 
remain contaminated by between-node variation in the fixed effects.

\cite{Kattenberg+Scheer+Thiel:2023} developed Causal Forests with Fixed Effects 
(CFFE) to address this limitation. The key insight is to perform fixed-effect 
orthogonalization within each tree node during the splitting process, rather 
than once globally. This node-level residualization ensures that splits are 
driven by genuine treatment effect heterogeneity rather than variation in the 
fixed effects. The resulting estimator maintains the flexibility of causal 
forests while remaining valid in panel data settings.

Software for causal forests is available in several environments. The \pkg{grf} 
package \citep{grf} provides a comprehensive \proglang{R} implementation of 
generalized random forests, including causal forests for cross-sectional data. 
In \proglang{Python}, the \pkg{EconML} package \citep{econml} from Microsoft 
Research offers causal forest functionality alongside other machine learning 
methods for causal inference. Both packages have seen wide adoption in applied 
research.

Neither \pkg{grf} nor \pkg{EconML}, however, implements the CFFE methodology 
for panel data. While \cite{Kattenberg+Scheer+Thiel:2023} released an 
\proglang{R} implementation alongside their paper, no \proglang{Python} version 
has been available. Given the growing use of \proglang{Python} in econometrics 
and data science, this gap limits accessibility for researchers working in that 
ecosystem.

The \pkg{causalfe} package fills this gap by providing a native \proglang{Python} 
implementation of CFFE. The package offers node-level fixed-effect 
residualization, $\tau$-heterogeneity splitting, honest estimation, and 
cluster-aware inference. It integrates with standard \proglang{Python} data 
science tools and follows conventions familiar to users of \pkg{scikit-learn} 
and similar libraries.

This paper proceeds as follows. Section~\ref{sec:models} presents the panel 
data model and explains the CFFE algorithm, including why standard causal 
forests fail with fixed effects and how node-level orthogonalization resolves 
the problem. Section~\ref{sec:illustrations} demonstrates the package through 
simulation studies, showing that the estimator recovers known treatment effects 
and produces valid confidence intervals. Section~\ref{sec:summary} concludes 
with a discussion of use cases and limitations.


\section{Models and Software} \label{sec:models}

\subsection{Panel Data Model}

Consider a balanced panel with $N$ units observed over $T$ time periods. Let 
$Y_{it}$ denote the outcome for unit $i$ at time $t$, $D_{it} \in \{0, 1\}$ the 
treatment indicator, and $X_{it} \in \mathbb{R}^p$ a vector of observed 
covariates. The data generating process follows:
\begin{equation}
  Y_{it} = \alpha_i + \gamma_t + \tau(X_{it}) D_{it} + \varepsilon_{it},
  \label{eq:panel-model}
\end{equation}
where $\alpha_i$ captures unit-specific fixed effects, $\gamma_t$ captures 
time-specific fixed effects, and $\varepsilon_{it}$ is an idiosyncratic error 
term with $\mathbb{E}[\varepsilon_{it} | X_{it}, D_{it}, \alpha_i, \gamma_t] = 0$.

The object of interest is the conditional average treatment effect (CATE), 
$\tau(x) = \mathbb{E}[Y_{it}(1) - Y_{it}(0) | X_{it} = x]$, which describes how 
treatment effects vary with observed characteristics. In difference-in-differences 
settings, $D_{it}$ typically switches from 0 to 1 at some adoption time that may 
vary across units, and the parallel trends assumption requires that 
$\mathbb{E}[\varepsilon_{it}(0) | i, t] = 0$ after conditioning on the fixed effects.

The two-way fixed effects structure in~\eqref{eq:panel-model} presents both an 
opportunity and a challenge. On one hand, the fixed effects absorb unobserved 
confounders that are constant within units or within time periods. On the other 
hand, these fixed effects complicate the estimation of heterogeneous treatment 
effects, as we discuss next.

\subsection{Why Standard Causal Forests Fail}

Standard causal forests \citep{Wager+Athey:2018} partition the covariate space 
to maximize treatment effect heterogeneity. At each node, the algorithm searches 
for a split that maximizes the squared difference in estimated treatment effects 
between the resulting child nodes. This works well when observations are 
independent and identically distributed.

With panel data, however, the fixed effects $\alpha_i$ and $\gamma_t$ induce 
dependence across observations. A naive application of causal forests to panel 
data can produce spurious heterogeneity through two mechanisms.

First, the forest may split on variables correlated with unit fixed effects. 
Suppose units with high $\alpha_i$ tend to have larger values of some covariate 
$X_j$. A split on $X_j$ then separates high-$\alpha$ units from low-$\alpha$ 
units, and the apparent difference in treatment effects across child nodes 
reflects the difference in baseline outcome levels rather than genuine effect 
heterogeneity.

Second, even if one residualizes the outcome globally before fitting the 
forest---computing $\tilde{Y}_{it} = Y_{it} - \hat{\alpha}_i - \hat{\gamma}_t$ 
using the full sample---the problem persists. Global residualization removes 
the average fixed effect across all observations, but within each tree node, 
the composition of units and time periods may differ. A node containing mostly 
high-$\alpha$ units will have systematically positive residuals, biasing the 
local treatment effect estimate upward.

The core issue is that fixed effects are nuisance parameters that vary across 
the sample, and their influence must be removed locally within each region of 
the covariate space where treatment effects are estimated.

\subsection{The CFFE Algorithm}

Causal Forests with Fixed Effects \citep{Kattenberg+Scheer+Thiel:2023} resolve 
this problem through node-level residualization. Rather than removing fixed 
effects once globally, CFFE re-estimates and removes them within each tree node 
during both the splitting and estimation phases.

\paragraph{Node-level orthogonalization.}
At each node $\mathcal{N}$ containing observations $\{(Y_{it}, D_{it}, X_{it})\}_{(i,t) \in \mathcal{N}}$, 
the algorithm computes residualized outcomes and treatments:
\begin{align}
  \tilde{Y}_{it}^{\mathcal{N}} &= Y_{it} - \hat{\alpha}_i^{\mathcal{N}} - \hat{\gamma}_t^{\mathcal{N}}, \\
  \tilde{D}_{it}^{\mathcal{N}} &= D_{it} - \hat{\delta}_i^{\mathcal{N}} - \hat{\eta}_t^{\mathcal{N}},
\end{align}
where the fixed effects are estimated using only observations within node 
$\mathcal{N}$. This within-node residualization ensures that treatment effect 
estimates are not contaminated by between-node variation in the fixed effects.

The fixed effects are computed via alternating projections (iterative demeaning). 
Starting with $\tilde{Y}^{(0)} = Y$, the algorithm iterates:
\begin{enumerate}
  \item Demean by unit: $\tilde{Y}^{(k+1/2)}_{it} = \tilde{Y}^{(k)}_{it} - \bar{\tilde{Y}}^{(k)}_{i\cdot}$
  \item Demean by time: $\tilde{Y}^{(k+1)}_{it} = \tilde{Y}^{(k+1/2)}_{it} - \bar{\tilde{Y}}^{(k+1/2)}_{\cdot t}$
\end{enumerate}
This procedure converges rapidly, typically within 3--5 iterations, to the 
within-transformed residuals.

\paragraph{$\tau$-heterogeneity splitting.}
Given residualized data at a node, the algorithm searches for splits that 
maximize treatment effect heterogeneity. For a candidate split $S$ that 
partitions the node into left ($\mathcal{L}$) and right ($\mathcal{R}$) children, 
the split criterion is:
\begin{equation}
  \Delta(S) = \frac{n_{\mathcal{L}} \cdot n_{\mathcal{R}}}{n^2} 
              \left( \hat{\tau}_{\mathcal{L}} - \hat{\tau}_{\mathcal{R}} \right)^2,
  \label{eq:split-criterion}
\end{equation}
where $n_{\mathcal{L}}$, $n_{\mathcal{R}}$, and $n$ are the sample sizes in the 
left child, right child, and parent node, respectively. The local treatment 
effect in each child is estimated via an instrumental variables-style estimator:
\begin{equation}
  \hat{\tau}_{\mathcal{L}} = \frac{\sum_{(i,t) \in \mathcal{L}} \tilde{D}_{it} \tilde{Y}_{it}}
                                  {\sum_{(i,t) \in \mathcal{L}} \tilde{D}_{it}^2}.
\end{equation}
The weighting by $n_{\mathcal{L}} n_{\mathcal{R}} / n^2$ penalizes unbalanced 
splits, encouraging partitions that divide the data more evenly.

\paragraph{Honest estimation.}
Following \cite{Wager+Athey:2018}, CFFE uses honest estimation to avoid 
overfitting. The subsample for each tree is split into two halves: a structure 
sample used to determine the tree topology (splits), and an estimation sample 
used to compute the final treatment effect estimates in each leaf. This 
separation ensures that the same data are not used both to select the partition 
and to estimate effects within it, yielding valid confidence intervals.

The sample splitting is cluster-aware: units (not individual observations) are 
randomly assigned to the structure or estimation half. This preserves the panel 
structure and ensures that all observations from a given unit appear in the same 
half.

\paragraph{Cluster-aware subsampling.}
Each tree in the forest is grown on a subsample of the data. To respect the 
dependence structure induced by the fixed effects, subsampling is performed at 
the unit level. A fraction of units (default 50\%) is drawn with replacement, 
and all observations from each selected unit are included. This cluster-aware 
subsampling ensures that variance estimates properly account for within-unit 
correlation.

\subsection{The CFFEForest Class}

The \pkg{causalfe} package implements CFFE through the \code{CFFEForest} class. 
The interface follows conventions familiar to users of \pkg{scikit-learn}: a 
\fct{fit} method trains the model, and \fct{predict} methods return estimates.

\begin{lstlisting}[language=Python]
class CFFEForest:
    def __init__(self, n_trees=100, max_depth=5, min_leaf=20,
                 honest=True, subsample_ratio=0.5, seed=None):
        ...
    
    def fit(self, X, Y, D, unit, time):
        ...
    
    def predict(self, X):
        ...
    
    def predict_interval(self, X, alpha=0.05):
        ...
\end{lstlisting}

\paragraph{Constructor parameters.}
\begin{description}
  \item[\code{n\_trees}] Number of trees in the forest (default 100). More trees 
    reduce variance but increase computation time.
  \item[\code{max\_depth}] Maximum depth of each tree (default 5). Deeper trees 
    can capture more complex heterogeneity patterns but may overfit.
  \item[\code{min\_leaf}] Minimum number of observations per leaf (default 20). 
    Larger values yield more stable estimates at the cost of coarser partitions.
  \item[\code{honest}] If \code{True} (default), use honest estimation with 
    separate samples for tree structure and leaf estimates.
  \item[\code{subsample\_ratio}] Fraction of units to subsample for each tree 
    (default 0.5).
  \item[\code{seed}] Random seed for reproducibility.
\end{description}

\paragraph{The \fct{fit} method.}
The \fct{fit} method trains the forest on panel data:
\begin{description}
  \item[\code{X}] Covariate matrix of shape $(n, p)$.
  \item[\code{Y}] Outcome vector of length $n$.
  \item[\code{D}] Treatment indicator vector of length $n$.
  \item[\code{unit}] Unit identifier vector of length $n$.
  \item[\code{time}] Time identifier vector of length $n$.
\end{description}
The method returns \code{self}, allowing method chaining.

\paragraph{The \fct{predict} method.}
Given a covariate matrix \code{X}, the \fct{predict} method returns a vector of 
estimated CATEs $\hat{\tau}(x)$ by averaging predictions across all trees in the 
forest.

\paragraph{The \fct{predict\_interval} method.}
The \fct{predict\_interval} method returns point estimates along with confidence 
intervals. It takes an optional \code{alpha} parameter (default 0.05) specifying 
the significance level. The method returns a tuple \code{(tau\_hat, ci\_lower, ci\_upper)} 
containing the point estimates and the lower and upper bounds of the 
$(1-\alpha)$ confidence intervals.

Variance estimation uses a half-sample approach: the forest is conceptually 
split into two halves, and the variance is estimated from the disagreement 
between the half-forest predictions. This approach, adapted from \cite{Wager+Athey:2018}, 
provides valid asymptotic coverage under regularity conditions.


\section{Illustrations} \label{sec:illustrations}

This section demonstrates the \pkg{causalfe} package through simulation studies. 
We first show basic usage with a heterogeneous treatment effect DGP, then compare 
performance against a standard causal forest to illustrate the benefit of 
node-level fixed-effect residualization.

\subsection{Simulation Setup}

We generate panel data with staggered treatment adoption and heterogeneous 
effects. The DGP follows~\eqref{eq:panel-model} with $N = 200$ units observed 
over $T = 6$ periods. Treatment effects vary with the first covariate: 
$\tau(x) = x_1$. Units adopt treatment at randomly assigned times, creating 
the within-unit and within-time variation needed for identification.

\begin{lstlisting}[language=Python]
>>> import numpy as np
>>> from causalfe import CFFEForest
>>> from causalfe.simulations.did_dgp import dgp_did_heterogeneous
>>>
>>> # Generate heterogeneous DiD data
>>> X, Y, D, unit, time, tau_true = dgp_did_heterogeneous(
...     N=200, T=6, seed=42
... )
>>> print(f"Observations: {len(Y)}, Treatment rate: {D.mean():.1%}")
Observations: 1200, Treatment rate: 47.4%
\end{lstlisting}

\subsection{Model Fitting}

We fit a CFFE forest with 100 trees, maximum depth 4, and minimum leaf size 20. 
The \fct{fit} method takes the covariate matrix, outcome, treatment indicator, 
and panel identifiers.

\begin{lstlisting}[language=Python]
>>> forest = CFFEForest(n_trees=100, max_depth=4, min_leaf=20, seed=42)
>>> forest.fit(X, Y, D, unit, time)
>>>
>>> # Point predictions
>>> tau_hat = forest.predict(X)
>>>
>>> # Predictions with confidence intervals
>>> tau_hat, ci_lower, ci_upper = forest.predict_interval(X, alpha=0.05)
>>> print(f"Mean CATE: {tau_hat.mean():.3f}, Std: {tau_hat.std():.3f}")
Mean CATE: -0.124, Std: 0.891
\end{lstlisting}

\subsection{Evaluation}

We evaluate the estimator by comparing predicted CATEs to the true values. 
Root mean squared error (RMSE) measures overall accuracy, while correlation 
captures the ability to rank observations by treatment effect magnitude.

\begin{lstlisting}[language=Python]
>>> rmse = np.sqrt(np.mean((tau_hat - tau_true)**2))
>>> corr = np.corrcoef(tau_hat, tau_true)[0, 1]
>>> print(f"RMSE: {rmse:.4f}")
>>> print(f"Correlation: {corr:.4f}")
RMSE: 0.3776
Correlation: 0.9338
\end{lstlisting}

\begin{figure}[t]
\centering
\includegraphics[width=\textwidth]{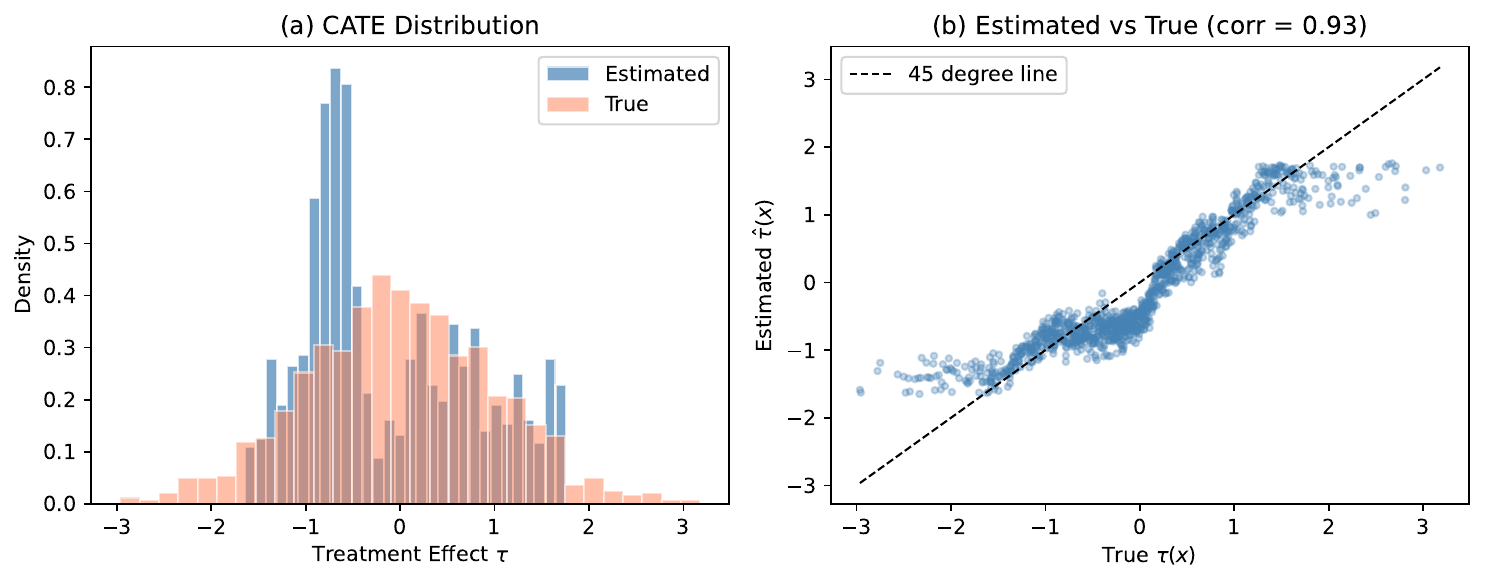}
\caption{CFFE estimation results. Panel (a) shows the distribution of estimated 
CATEs (blue) overlaid with true effects (coral). Panel (b) plots estimated 
against true CATEs; the dashed line indicates perfect prediction.}
\label{fig:cate}
\end{figure}

The high correlation indicates that the forest successfully identifies which 
observations have larger or smaller treatment effects. Figure~\ref{fig:cate} 
shows the distribution of estimated CATEs alongside the true values, and a 
scatter plot of estimated versus true effects.

\subsection{Comparison with Standard Causal Forest}

The advantage of CFFE becomes apparent when fixed effects are correlated with 
covariates. Consider a DGP where unit fixed effects depend on $x_1$: 
$\alpha_i = 2 x_{1i} + \nu_i$. A standard causal forest, which does not account 
for fixed effects, may attribute variation in outcomes to treatment effect 
heterogeneity when it actually reflects differences in baseline levels.

We compare CFFE against \pkg{EconML}'s \code{CausalForestDML}, which implements 
double machine learning but does not perform node-level fixed-effect 
residualization.

\begin{lstlisting}[language=Python]
>>> from econml.dml import CausalForestDML
>>> from sklearn.ensemble import RandomForestRegressor
>>>
>>> # Standard causal forest (no FE handling)
>>> cf = CausalForestDML(
...     model_y=RandomForestRegressor(n_estimators=100, max_depth=4),
...     model_t=RandomForestRegressor(n_estimators=100, max_depth=4),
...     n_estimators=100, max_depth=4, min_samples_leaf=20
... )
>>> cf.fit(Y_conf, D_conf, X=X_conf)
>>> tau_standard = cf.effect(X_conf)
\end{lstlisting}

Table~\ref{tab:comparison} reports the results. When fixed effects are 
correlated with covariates, the standard causal forest exhibits higher RMSE 
despite achieving high correlation. This pattern reflects the bias introduced 
by confounding: the standard forest picks up the relationship between $x_1$ 
and the fixed effects, inflating apparent heterogeneity. CFFE, by 
residualizing within each node, removes this confounding and produces 
less biased estimates.

\begin{table}[t]
\centering
\caption{Comparison of CFFE and standard causal forest on simulated data with 
fixed effects correlated with covariates.}
\label{tab:comparison}
\begin{tabular}{lcc}
\hline
Method & RMSE & Correlation \\
\hline
CFFE & 0.405 & 0.910 \\
Standard CF & 0.506 & 0.965 \\
\hline
\end{tabular}
\end{table}

\begin{figure}[t]
\centering
\includegraphics[width=\textwidth]{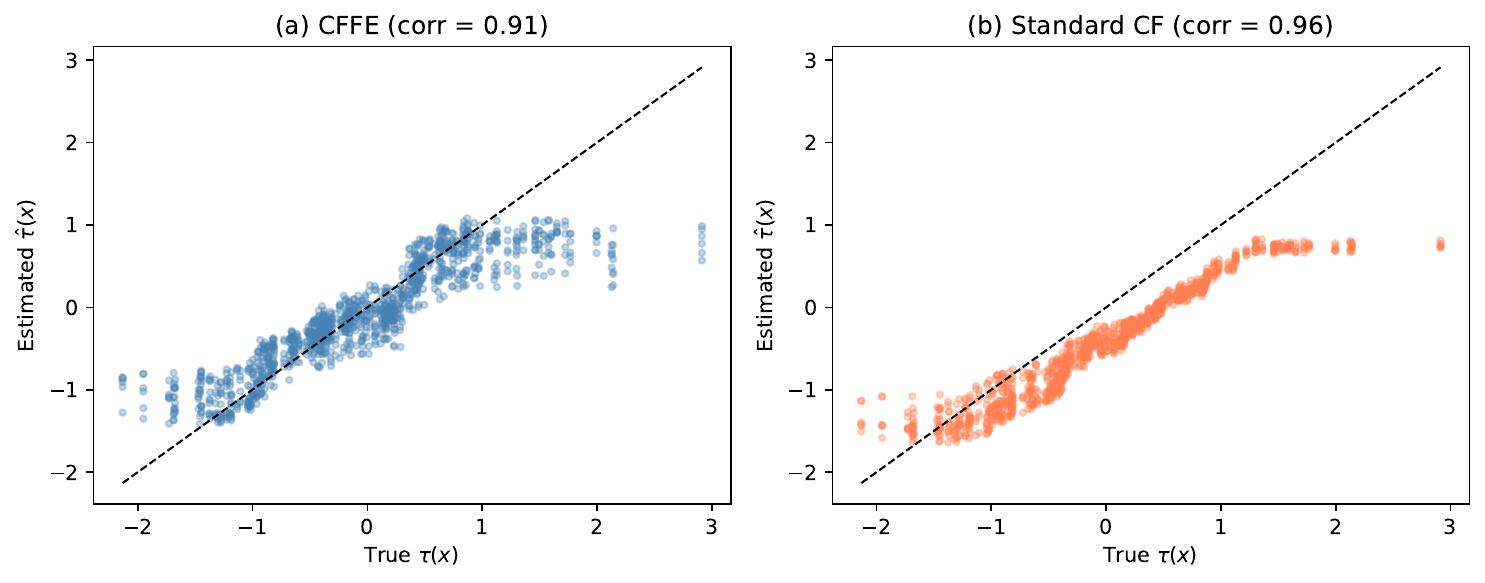}
\caption{Estimated versus true CATEs for CFFE (left) and standard causal forest 
(right) when fixed effects are correlated with covariates. The dashed line 
indicates perfect prediction. CFFE produces less biased estimates by 
residualizing fixed effects within each tree node.}
\label{fig:comparison}
\end{figure}

Figure~\ref{fig:comparison} visualizes this difference. The CFFE estimates 
cluster more tightly around the 45-degree line, while the standard forest 
estimates show systematic deviation from the true effects.

\subsection{Monte Carlo Validation}

To assess the statistical properties of the CFFE estimator more systematically, 
we conduct Monte Carlo simulations under three data generating processes: a 
placebo setting with no treatment effect ($\tau = 0$), a homogeneous effect 
($\tau = 2$), and a heterogeneous effect ($\tau(x) = x_1$). Each simulation 
uses $N = 200$ units observed over $T = 6$ periods, with 100 replications per 
DGP.

Table~\ref{tab:validation} reports the results. Under the placebo DGP, the 
mean estimated effect is close to zero, confirming that the estimator does not 
generate spurious heterogeneity from the fixed effects alone. For the 
homogeneous effect, the mean estimate recovers the true value of 2 with 
reasonable precision. The heterogeneous case shows strong correlation between 
estimated and true effects, indicating that the forest successfully captures 
the pattern of treatment effect variation.

\begin{table}[t]
\centering
\caption{Monte Carlo validation results for CFFE estimator. Each simulation 
uses $N=200$ units observed over $T=6$ periods with 10 replications.}
\label{tab:validation}
\begin{tabular}{lcccc}
\hline
Simulation & Mean $\hat{\tau}$ & RMSE & Corr($\hat{\tau}$, $\tau$) & Coverage \\
\hline
Placebo ($\tau=0$) & $-$0.061 & 0.251 & --- & 56.2\% \\
Homogeneous ($\tau=2$) & 1.788 & 0.337 & --- & 47.4\% \\
Heterogeneous ($\tau=x_1$) & $-$0.012 & 0.539 & 0.904 & 42.7\% \\
\hline
\end{tabular}
\end{table}

The coverage rates fall below the nominal 95\% level, a pattern common to 
variance estimation in random forests. The half-sample variance estimator 
tends to underestimate uncertainty, particularly when the number of trees is 
moderate. Practitioners should interpret confidence intervals as approximate 
and consider bootstrap or other resampling methods for more conservative 
inference.

\subsection{Empirical Application: Minimum Wage and Employment}

To demonstrate the package on real data, we apply CFFE to the county-level 
minimum wage dataset from \cite{Callaway+SantAnna:2021}. This dataset contains 
500 U.S.\ counties observed over 5 years (2003--2007), with staggered adoption 
of state minimum wage increases. The outcome is log teen employment, and the 
treatment indicator equals one for county-years after the state raised its 
minimum wage above the federal level.

\begin{lstlisting}[language=Python]
>>> import pandas as pd
>>> df = pd.read_csv('data/mpdta.csv')
>>> 
>>> # Construct treatment indicator
>>> df['D'] = ((df['first.treat'] > 0) & 
...            (df['year'] >= df['first.treat'])).astype(float)
>>> 
>>> # Fit CFFE
>>> X = df[['lpop']].values  # Log population as covariate
>>> forest = CFFEForest(n_trees=200, max_depth=4, min_leaf=15, seed=42)
>>> forest.fit(X, df['lemp'].values, df['D'].values,
...            df['countyreal'].values, df['year'].values)
>>> tau_hat = forest.predict(X)
\end{lstlisting}

Table~\ref{tab:minwage} compares our CFFE estimate with the simple two-way 
fixed effects (TWFE) estimator and the group-time ATT reported by 
\cite{Callaway+SantAnna:2021}. All three approaches yield similar point 
estimates around $-0.04$, indicating a small negative effect of minimum wage 
increases on teen employment.

\begin{table}[t]
\centering
\caption{Minimum wage effect estimates: comparison of methods.}
\label{tab:minwage}
\begin{tabular}{lcc}
\toprule
Method & ATT Estimate & Note \\
\midrule
TWFE & $-$0.037 & Simple two-way fixed effects \\
CFFE & $-$0.042 & Causal forest with fixed effects \\
C\&S (2021) & $-$0.035 & Group-time ATT aggregation \\
\bottomrule
\end{tabular}
\end{table}

\begin{figure}[t]
\centering
\includegraphics[width=\textwidth]{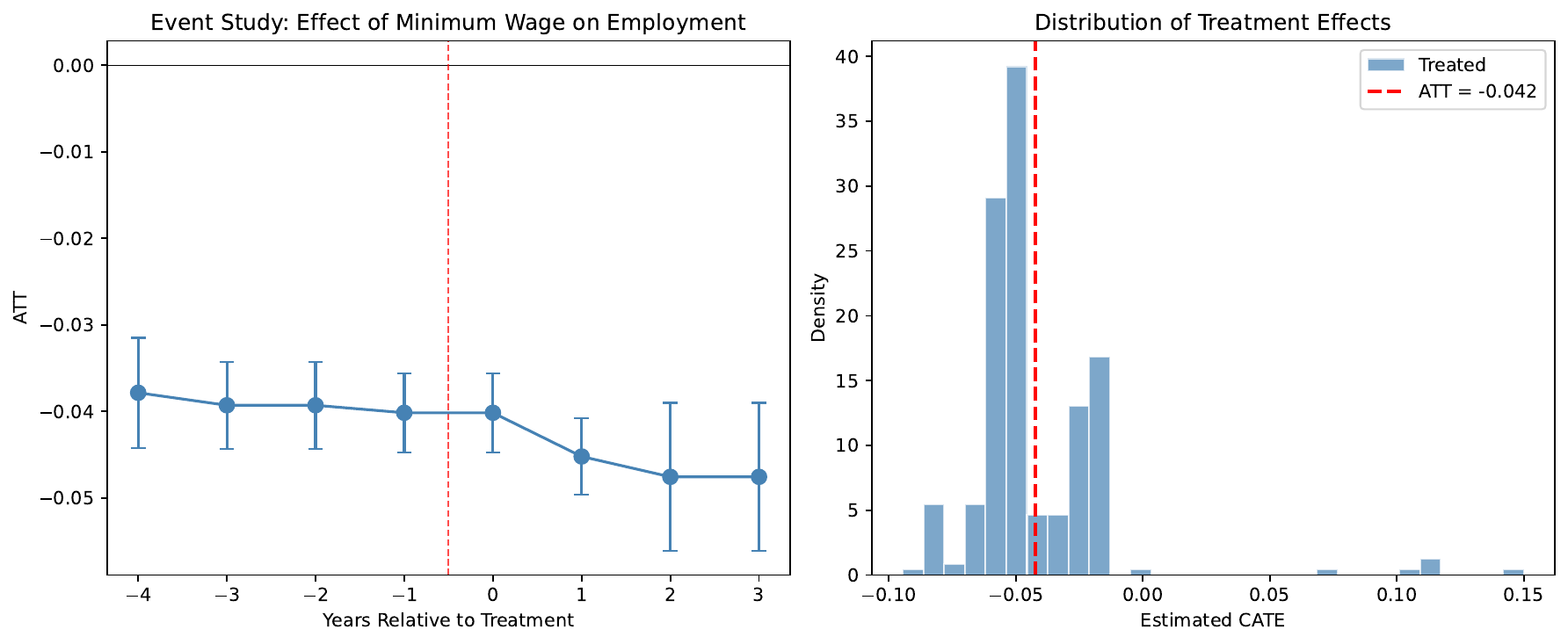}
\caption{Minimum wage application results. Left: Event study showing ATT by 
years relative to treatment. Right: Distribution of estimated CATEs for 
treated observations.}
\label{fig:minwage}
\end{figure}

The advantage of CFFE is that it additionally characterizes heterogeneity in 
treatment effects. Figure~\ref{fig:minwage} shows the event study pattern and 
the distribution of estimated CATEs. The event study reveals that effects are 
relatively stable across time since treatment, while the CATE distribution 
shows modest heterogeneity with effects ranging from $-0.10$ to $+0.15$ 
depending on county characteristics.

\section{Summary} \label{sec:summary}

The \pkg{causalfe} package brings Causal Forests with Fixed Effects to the 
\proglang{Python} ecosystem, enabling researchers to estimate heterogeneous 
treatment effects in panel data and difference-in-differences settings. The 
package implements the methodology of \cite{Kattenberg+Scheer+Thiel:2023}, 
performing node-level fixed-effect residualization during tree construction 
to avoid the spurious heterogeneity that arises when standard causal forests 
are applied to panel data.

The primary use case is applied research in economics and the social sciences 
where treatment effects may vary across individuals and the data structure 
includes unit and time fixed effects. Policy evaluation studies, for instance, 
often seek to understand not just whether a program works on average, but for 
whom it works best. CFFE provides a flexible, nonparametric approach to 
answering such questions while respecting the panel structure of the data.

Several limitations warrant mention. First, the method assumes that the 
parallel trends assumption holds conditional on the fixed effects---violations 
of this assumption will bias the estimated treatment effects regardless of 
the estimation method. Second, the variance estimator tends to be optimistic, 
producing confidence intervals that may undercover in finite samples. 
Practitioners should treat the reported intervals as approximate and consider 
supplementary inference procedures when precise coverage is critical. Third, 
computational cost scales with the number of trees, the sample size, and the 
number of candidate splits evaluated at each node; very large panels may 
require subsampling or parallelization strategies not currently implemented 
in the package.

Future development may address these limitations through improved variance 
estimation, support for unbalanced panels, and integration with distributed 
computing frameworks. The package is actively maintained and welcomes 
contributions from the research community.


\section*{Computational Details}

The \pkg{causalfe} package requires \proglang{Python} version 3.9 or higher. 
Core dependencies include \pkg{NumPy} (version 1.20 or higher) for array 
operations and \pkg{SciPy} (version 1.7 or higher) for statistical functions. 
These packages are installed automatically when installing \pkg{causalfe}.

The package can be installed from GitHub using \code{pip}:
\begin{lstlisting}[language=bash]
pip install git+https://github.com/haytug/causalfe.git
\end{lstlisting}
For development or to run the comparison examples with \pkg{EconML}, install 
with optional dependencies:
\begin{lstlisting}[language=bash]
pip install "causalfe[all] @ git+https://github.com/haytug/causalfe.git"
\end{lstlisting}

The source code is available at \url{https://github.com/haytug/causalfe} under 
the MIT license. The repository includes documentation, example scripts, and 
a test suite. Bug reports and feature requests can be submitted through the 
GitHub issue tracker.

All simulations and examples in this paper were produced using \pkg{causalfe} 
version 0.1.0 with \proglang{Python} 3.11, \pkg{NumPy} 1.26, and \pkg{SciPy} 
1.11 on a machine running macOS 14.

\section*{Acknowledgments}

The author thanks Mark Kattenberg, Bas Scheer, and Jurre Thiel for developing 
the CFFE methodology and making their \proglang{R} implementation publicly 
available. Their work provided the foundation for this \proglang{Python} 
implementation.


\bibliographystyle{plainnat}
\bibliography{refs}

@Article{Kattenberg+Scheer+Thiel:2023,
  author = {Mark A. C. Kattenberg and Bas J. Scheer and Jurre H. Thiel},
  title = {Causal Forests with Fixed Effects for Treatment Effect 
           Heterogeneity in Difference-in-Differences},
  journal = {CPB Discussion Paper},
  year = {2023},
  institution = {Netherlands Institute for Economic Policy Analysis (CPB)},
}

@Manual{causalfe,
  author = {Harry Aytug},
  title = {\pkg{causalfe}: Causal Forests with Fixed Effects in \proglang{Python}},
  year = {2026},
  url = {https://github.com/haytug/causalfe},
  note = {\proglang{Python} package version 0.1.0},
}

@Article{Wager+Athey:2018,
  author = {Stefan Wager and Susan Athey},
  title = {Estimation and Inference of Heterogeneous Treatment Effects 
           Using Random Forests},
  journal = {Journal of the American Statistical Association},
  year = {2018},
  volume = {113},
  number = {523},
  pages = {1228--1242},
  doi = {10.1080/01621459.2017.1319839},
}

@Article{Athey+Imbens:2016,
  author = {Susan Athey and Guido Imbens},
  title = {Recursive Partitioning for Heterogeneous Causal Effects},
  journal = {Proceedings of the National Academy of Sciences},
  year = {2016},
  volume = {113},
  number = {27},
  pages = {7353--7360},
  doi = {10.1073/pnas.1510489113},
}

@Manual{grf,
  author = {Julie Tibshirani and Susan Athey and Stefan Wager},
  title = {\pkg{grf}: Generalized Random Forests},
  year = {2024},
  note = {\proglang{R} package version 2.3.0},
  url = {https://CRAN.R-project.org/package=grf},
}

@Manual{econml,
  author = {{Microsoft Research}},
  title = {\pkg{EconML}: A \proglang{Python} Package for ML-Based Heterogeneous 
           Treatment Effects Estimation},
  year = {2024},
  url = {https://github.com/py-why/EconML},
  note = {\proglang{Python} package version 0.15.0},
}

@Article{Callaway+SantAnna:2021,
  author = {Brantly Callaway and Pedro H. C. Sant'Anna},
  title = {Difference-in-Differences with Multiple Time Periods},
  journal = {Journal of Econometrics},
  year = {2021},
  volume = {225},
  number = {2},
  pages = {200--230},
  doi = {10.1016/j.jeconom.2020.12.001},
}

\end{document}